\documentclass{article}
\usepackage[utf8]{inputenc}
\usepackage[sort,compress,numbers]{natbib}
\usepackage{graphicx,xcolor}
\usepackage{subfigure}
\usepackage{amsmath}

\newcommand{\bea}{\begin{eqnarray}}
\newcommand{\eea}{\end{eqnarray}}

\newcommand{\braket}[2]{\left< #1 \vphantom{#2} \right|
 \left. #2 \vphantom{#1} \right>} 
\newcommand{\matrixel}[3]{\left< #1 \vphantom{#2#3} \right|
 #2 \left| #3 \vphantom{#1#2} \right>} 
 \newcommand{\ev}[1]{\left\langle #1 \right\rangle} 

\begin{document}
\markboth{ Ahmed Al-Jamel}{}

%
%

%
%

\title{\textbf{ WKB Approximation with Conformable Operator} }

\author{Mohamed.Al-Masaeed, Eqab.M.Rabei and Ahmed Al-Jamel\\
Physics Department, Faculty of Science, Al al-Bayt University,\\ P.O. Box 130040, Mafraq 25113, Jordan\\moh.almssaeed@gmail.com\\eqabrabei@gmail.com\\
aaljamel@aabu.edu.jo, aaljamel@gmail.com}

\maketitle


\begin{abstract}
In this paper, the WKB method is  extended to be applicable  for conformable Hamiltonian  systems where  the concept of conformable operator  with fractional order $\alpha$ is used. The WKB approximation for the $\alpha$-wavefunction is derived  when the potential is slowly varying in space. The paper is furnished with some illustrative examples to demonstrate the method. The quantities of the conformable form are found to be in exact agreement with traditional quantities when $\alpha=1$. 
\\

\end{abstract}
\section{Introduction}
Approximation methods in quantum mechanics, such as the variational method, WKB approximation, and perturbation theory are important tools. Each of these methods has its area of applicability that depend on the nature of the physical system under investigation. The Wentzel-Kramers-Brillouin (WKB) method is useful, 
with potentials that vary slowly; that is, potentials that remain almost constant over a potential that varies region of the de Broglie Wavelength order. This property, in the case of classical systems, since a classical system's wavelength reaches zero, the WKB is always fulfilled. It is also possible to consider the approach as a semi-classical approximation\cite{griffithsIntroductionQuantumMechanics2018,zettili2003quantum}.

This method is named after Wentzel, Kramers, and Brillouin physicists, all of whom invented it in 1926 \cite{kramers1926wellenmechanik,wentzel1926verallgemeinerung}. Shortly before that, Harold Jeffreys, a mathematician, had already developed a general method of approximating linear ordinary differential equation solutions in 1923, but the other three were unaware of his work. It is thus commonly referred to today as the WKB or WKBJ approximation \cite{jeffreys1925certain}. Rabie et.al.\cite{rabei2002quantization} were able to quantize constrained systems by using the WKB approximation method, where, after the quantization process, the constraints became conditions to be met in the semiclassical limit on the wave function. Besides, many researchers have used this method to create a quantization process of physical systems\cite{rabei2003quantization,nawafleh2005quantization,rabei2005quantization,hasan2004quantization}.

The fractional derivative is as ancient as calculus. In 1695, L'Hospital asked what it meant $\frac{d^n f}{dx^n}$ if $n=\frac{1}{2}$. Since then,
researchers have attempted to describe a fractional derivative. Most of them used an integral form to define the fractional derivative. there many different definitions for fractional derivative,Riemann-Liouville, Caputo, Riesz, Weyl, Grünwald, Riesz-Caputo, Chen, and Hadamard \cite{caputoLinearModelsDissipation1967,oldhamFractionalCalculusTheory1974,millerIntroductionFractionalIntegrals1993,kilbasTheoryApplicationsFractional,hilferApplicationsFractionalCalculus2000,podlubnyFractionalDifferentialEquations1998,klimekLagrangeanHamiltonianFractional2002,chenInvestigationFractionalFractal2010,heNewFractalDerivation2011,gorenfloEssentialsFractionalCalculus2000,kimeuFractionalCalculusDefinitions2009}. 
Two of which are among the most common, Riemann-Liouville and Caputo.

 Physicists have investigated WKB approximation process of quantization for the fractional Hamiltonian \cite{rabei2009fractional,sayevandModificationBackslashMathsf2017,rabei2018quantization}. A new concept of  derivative, called the conformable  derivative is given by Khalil et.al.\cite{khalil2014new}, was introduced a few years ago. This description satisfies the conventional derivative's standard properties. With the emergence of many definitions, the classification of these definitions has started according to the non-locality (Time-based nonlocality is typically called memory) characteristic to local  as the conformable  derivative, the M-fractional derivative, the alternative fractional derivative, the local fractional derivative, and the Caputo-Fabrizio fractional derivatives with exponential kernels derivative \cite{khalil2014new,katugampola2014new,sousa2017new,caputo2015new,kolwankar1996fractional,kolwankar2013local,kolwankar1997holder}, and nonlocal fractional derivative as the Riemann-Liouville, Caputo, Hadamard, Marchaud fractional derivatives\cite{caputoLinearModelsDissipation1967,oldhamFractionalCalculusTheory1974,millerIntroductionFractionalIntegrals1993,kilbasTheoryApplicationsFractional,hilferApplicationsFractionalCalculus2000,podlubnyFractionalDifferentialEquations1998}. Theoretically, the local  derivative is much easier to handle and also obeys certain conventional properties that can not be met by the other  derivatives, and it seems to satisfy all the requirements of the standard derivative. For example, the chain law. Thus, the local operator derivative appropriate to extend the WKB approximation. 
 
 Recently in ref \cite{chung2020effect}. Using the conformable operator calculus, the deformation of ordinary quantum mechanics is discussed. The $\alpha$ Hamiltonian operator is suggested and the conformable Schrodinger  equation is constructed. Based on that, we  quantized  fractional harmonic oscillator using creation and annihilation operators in ref \cite{AlMasaeedRabeiAlJamelBaleanu+2021+395+401}. In addition , the conformable calculus was used to extend the perturbation theory to quantum systems containing a conformable derivative of fractional order $\alpha$ in a recent paper \cite{al2021extension}. In this work, we would like to extend the WKB approximation to include the conformable derivative  of $\alpha$ order.

\section{Conformable  Derivative}
\textbf{Definition 2.1.} Given a function $f\in [0,\infty) \to {R}$. The conformable  derivative of $f$ with order $\alpha$ is defined by \cite{khalil2014new}
\begin{equation}
\label{conformable}
T_\alpha(f)(t)=\lim_{\epsilon \to 0}\frac{f(t+\epsilon t^{1-\alpha})-f(t)}{\epsilon}
\end{equation}
for all $t>0$, $\alpha\in (0.1)$
In this paper, we adopt $D^\alpha f$ to denote the conformable  derivative (CD) of $f$ of order $\alpha$. The properties of this derivative is discussed in detail in Ref \cite{khalil2014new}.\\
 \textbf{Definition 2.2.}.$I_\alpha^a (f)(t)=I_1^a (t^{\alpha-1}f)= \int_a^t \frac{f(x)}{x^{1-\alpha}} dx$ where the integral is the usual Riemann improper integral and $\alpha \in (0,1)$. The conformable integral and the conformable derivative obey the following relations
\begin{equation}
 D^\alpha I_\alpha^a f(x)=f(x) 
\end{equation}
\begin{equation}
 I_\alpha^a  D^\alpha f(x)=f(x) -f(a)
\end{equation}
 To read more about conformable  derivative, its properties, and its applications, we refer you to \cite{abdeljawad2015conformable,hammad2014abel,atangana2015new,hammad2021analytical,khalil2014conformable,khalil2019geometric,singh2018numerical,al2019search,arqub2020fuzzy,chung2021new,mozaffari2018investigation,atraoui2021existence,zhao2017general}.
 \section{The conformable quantum mechanics}
In terms the conformable  derivative, the postulates of conformable quantum mechanics are presented in \cite{mozaffari2018conformable}. The momentum and the coordinate are defined as 
\begin{equation}
\label{coor xp}
\hat{x}_\alpha=x,~~~ \hat{p}_\alpha=-i\hbar_\alpha^\alpha D^\alpha_x,
\end{equation}
and $\hbar_\alpha^\alpha=\frac{h}{(2\pi)^{\frac{1}{\alpha}}}$. In the Hilbert space, the inner product is given by 
\begin{equation}
\label{inner}
\braket{f}{g}=\int_{-\infty}^\infty g^*(x) f(x) |x|^{\alpha-1} dx
\end{equation}
Thus, the expectation value of an observable $A$ for a system in the state $\psi(x,t)$ is given by 
\begin{equation}
\begin{aligned}
\label{expectation}
\ev{A}&=\matrixel{\psi(x,t)}{A}{\psi(x,t)}\\&=\int_{-\infty}^\infty \psi^*(x,t) A\psi(x,t) |x|^{\alpha-1} dx.
\end{aligned}
\end{equation}
In conformable quantum mechanics, the commutator of the $\alpha-$ position operator  $\hat{x}_\alpha$ and $\alpha-$ momentum operator $\hat{p}_\alpha$ is given as \cite{mozaffari2018conformable}: 
\begin{equation}
\label{commutator}
[\hat{x}_\alpha,\hat{p}_\alpha]=i\hbar_\alpha^\alpha |x|^{1-\alpha}.
\end{equation}
Following,Chung et.al \cite{chung2020effect},the continuity equation is found to be as 
\begin{equation}
D^\alpha_t \rho_\alpha(x,t)+D^\alpha_x j_\alpha(x,t)=0
\end{equation}
where the $\alpha-$probability density $\rho_\alpha(x,t)$ is given as 
\begin{equation}
\rho_\alpha(x,t)=\psi^* \psi,
\end{equation}
and the $\alpha-$probability flux $j_\alpha(x,t)$ is given as 
\begin{equation}
j_\alpha(x,t)=\frac{\hbar_\alpha^\alpha}{2im^\alpha}(\psi^*D^\alpha_x\psi-\psi D^\alpha_x \psi^*).
\end{equation}
 \section{Conformable WKB Approximation }
To introduce the idea behind this approximation using conformable  derivative, we first consider the Schrodinger equation as
\begin{equation}
\label{tise wkb}
\frac{\hat{p}_\alpha^2}{2m^\alpha} \psi_\alpha(x,t)=(E^\alpha -V_\alpha(\hat{x}_\alpha))\psi_\alpha(x,t).
\end{equation}
We want to study this conformable Schrodinger equation in two cases.\\
\textbf{ First case}. For constant potential $V_\alpha(x) = V^\alpha$. 
 Then, we can rewrite eq.(\ref{tise wkb}) as  
\begin{equation}
 \label{tise f1}
\hat{p}_\alpha \psi_\alpha(x) = \pm \sqrt{2m^\alpha(E^\alpha -V^\alpha)}\psi_\alpha(x),
 \end{equation}
and using eq.(\ref{coor xp}), we have
\begin{equation}
 D^\alpha_x \psi_\alpha(x)= \pm i k \psi_\alpha(x),
\end{equation}
where $k=\frac{\sqrt{2m^\alpha(E^\alpha -V^\alpha)}}{\hbar_\alpha^\alpha} $. The form of the solution for this conformable differential equation is:
\begin{equation}
\label{psi k}
\psi_\alpha(x)= A \exp{\left( \pm i k \frac{x^\alpha}{\alpha}\right)}.
\end{equation}
The conformable wave function differs due to  two states: classical case and quantum case\\
1- Classical case: when  $E^\alpha > V^\alpha \to k=\frac{\sqrt{2m^\alpha(E^\alpha -V^\alpha)}}{\hbar_\alpha^\alpha}$, where $k$ is real number thus, the solution for conformable Schrodinger equation in this case is given by eq.(\ref{psi k})\\
2- Quantum case: when $E^\alpha < V^\alpha \to k=i q =i\frac{\sqrt{2m^\alpha(V^\alpha -E^\alpha )}}{\hbar_\alpha^\alpha}$, where q is a real number and k is imaginary, thus, the solution for conformable Schrodinger equation, in this case, is given by 
\begin{equation} 
\label{psi q}
\psi_\alpha(x)= A \exp{\left(\pm q \frac{x^\alpha}{\alpha} \right)}.
\end{equation}
The conformable wave function $ \psi$  is increasing if the sign is positive, and is decreasing if the sign is negative.\\
\textbf{Second case}. For variable potential $V_\alpha(x) $ where the potential is slowly varying. We  study this conformable wave function in two cases:\\
1- Classical case; when  $E^\alpha > V_\alpha(x)$, since there is potential  that varies slowly. We expect the solution of conformable Schrodinger equation will be in the form ,
\begin{equation}
\label{c1}
\psi_\alpha(x)= A(x) \exp{(\pm i \phi(x))},
\end{equation}
where $A(x)$ is the amplitude  and $\phi(x)$ is the phase, which both depends on $x$.\\
Thus, we want to calculate $A(x) $ and $\phi(x)$, using 
\begin{equation}
D^\alpha_x D^\alpha_x \psi(x) =-\frac{\hat{p}_\alpha^2}{\hbar_\alpha^{2\alpha}}\psi(x).
\end{equation}
After substituting eq.(\ref{c1}), we have two parts:\\
- equating real part, we get
\begin{equation}
\label{real part}
D^\alpha_x D^\alpha_x A(x)  - A(x) (D^\alpha_x \phi(x))^2=-\frac{\hat{p}_\alpha^2}{\hbar_\alpha^{2\alpha}}A(x).
\end{equation}
- and equating imaginary part, we obtain
\begin{equation}
\label{imaginary part}
2D^\alpha_x A(x) D^\alpha_x \phi(x) +A(x) D^\alpha_x D^\alpha_x \phi(x)=0.
\end{equation}
eq.(\ref{real part}) leads to 
\begin{equation}
\label{phi fun}
\phi(x)= \pm \frac{1}{\hbar^{\alpha}_\alpha}\int p_\alpha (x) d^\alpha x,
\end{equation}
where $D^\alpha_x D^\alpha_x A(x) \approx 0 $, because $A(x)$ varies slowly, so that the $D^\alpha_x D^\alpha_x A(x)$ term is negligible. More precisely, we assume that $ \frac{D^\alpha_x D^\alpha_x A(x)}{A(x)}$ is much less than both $(D^\alpha_x \phi(x))^2$ and $\frac{\hat{p}_\alpha^2}{\hbar_\alpha^{2\alpha}}$. In that case we can drop the first term of the lift side of eq.\eqref{real part}
But eq.(\ref{imaginary part})  can be rewritten  as 
\begin{equation}
\label{21}
D^\alpha_x (A^2(x)  D^\alpha_x \phi(x) ) =0,
\end{equation}
where this equation is easily solved 
\begin{equation}
A^2(x)  D^\alpha_x \phi(x)  = c^2, or \quad A(x)=\frac{c}{\sqrt{D^\alpha_x \phi(x)}}.
\end{equation}
Thus, after substituting $D^\alpha_x \phi(x)=\frac{\hat{p}_\alpha}{\hbar_\alpha^{\alpha}}$, we obtain 
\begin{equation}
\label{A}
A(x)=\frac{C}{\sqrt{p^\alpha (x)}},
\end{equation}
where $C = c \sqrt{\hbar_\alpha^{\alpha}}$. Then substituting eqs.(\ref{phi fun})and(\ref{A}) in eq.(\ref{c1}), we obtain 
\begin{equation}
\label{psi general}
\psi_\alpha(x)= \frac{C}{\sqrt{p^\alpha (x)}} \exp { \left( \pm  \frac{i}{\hbar^{\alpha}_\alpha}\int p^\alpha (x) d^\alpha x \right)},
\end{equation}
which can be written in the following form
\begin{equation}
\begin{aligned}
\label{psi sin}
\psi_\alpha(x) &= \frac{C_1}{\sqrt{p^\alpha (x)}} \sin{\left( \frac{1}{\hbar^{\alpha}_\alpha}\int p^\alpha (x) d^\alpha x\right)}\\&+\frac{C_2}{\sqrt{p^\alpha (x)}} \cos{\left(  \frac{1}{\hbar^{\alpha}_\alpha}\int p^\alpha (x) d^\alpha x \right)},
\end{aligned}
\end{equation}
2- Quantum case (Tunneling): when $E^\alpha < V_\alpha(x) $, so, the solution of fractional Schrodinger equation is given by 
\begin{equation}
\label{psi general}
\psi_\alpha(x)= \frac{C}{\sqrt{|p^\alpha (x)|}} \exp{\left(\pm  \frac{1}{\hbar^{\alpha}_\alpha}\int |p^\alpha(x)| d^\alpha x \right)},
\end{equation}
where $p^\alpha (x)=i\sqrt{2m^\alpha(V_\alpha(x) -E^\alpha )}$.
We can estimate  the probability of transmission in conformable form using \cite{griffithsIntroductionQuantumMechanics2018}
\begin{equation}
\label{tran}
T_\alpha=\exp{\left(-2 \gamma_\alpha \right)},
\end{equation}
where $\gamma_\alpha$ is called the $\alpha-$Gamow factor which is defined as 
\begin{equation}
\label{gamow}
\gamma_\alpha= \frac{1}{\hbar^{\alpha}_\alpha}\int_0^a |p^\alpha| (x) d^\alpha x
\end{equation}
\section{Alternative approach }
We present here an Alternative approach by Making use of the relation between the conformable wave function and the conformable Hamilton's principal function $S$. The exponential solution of the conformable Schrodinger equation can be written as \cite{griffithsIntroductionQuantumMechanics2018}
\begin{equation}
\label{psi S}
\psi_\alpha(x)= A \exp{\left(  \frac{i}{\hbar^{\alpha}_\alpha} S(x) \right)}.
\end{equation}
The conformable Schrodinger equation for free particle can be written in the form 
\begin{equation}
\label{sch 2}
D^\alpha_x D^\alpha_x \psi(x) +\frac{p^{2\alpha}}{\hbar_\alpha^{2\alpha}}\psi(x)=0.
\end{equation}
Substituting eq.(\ref{psi S}) in this equation, we have
\begin{equation}
\label{sch S1}
[(D^\alpha_x S(x))^2-i \hbar^{\alpha}_\alpha D^\alpha_x D^\alpha_x S(x) -p^{2\alpha}  ]\psi(x) =0.
\end{equation}
By writing $S(x)$ as power series of $\hbar^{\alpha}_\alpha$
\begin{equation}
\label{S series}
S(x)=S_0(x)+S_1(x) \hbar^{\alpha}_\alpha +S_2(x) \hbar^{2\alpha}_\alpha+\dots,
\end{equation}
and substituting in eq.(\ref{sch S1}), we obtain
\begin{equation}
\label{S0}
\hbar^{0}_\alpha \to (D^\alpha_x S_0(x))^2=p^{2\alpha}, 
\end{equation}
\begin{equation}
\label{S1}
\hbar^{\alpha}_\alpha \to 2D^\alpha_x S_0(x)D^\alpha_x S_1(x)-i D^\alpha_x D^\alpha_x S_0(x)= 0.
\end{equation}
Thus, from these equations we have 
\begin{equation}
\label{S0 def}
 S_0(x)= \pm \int p^{\alpha}  d^\alpha x, 
 \end{equation}
 \begin{equation} \label{S1}
  S_1(x) = \frac{i}{2} \ln{p^{\alpha} }.
\end{equation}
So, after substituting these equations in eq.(\ref{psi S}) we obtain
\begin{equation}
\label{psi S2}
\psi_\alpha(x)= \frac{A}{\sqrt{p^{\alpha} }} \exp{\left(  \pm \frac{i}{\hbar^{\alpha}_\alpha} \int p^{\alpha}  d^\alpha x  \right)}.
\end{equation}
This result in classical case when $E^\alpha > V_\alpha(x) $, and for quantum case when $E^\alpha < V_\alpha(x) $ , so, eq.(\ref{psi S2}) becomes 
\begin{equation}
\psi_\alpha(x)= \frac{A}{\sqrt{|p^{\alpha}| }} \exp{\left(  \pm \frac{1}{\hbar^{\alpha}_\alpha} \int |p^{\alpha}|d^\alpha x \right) }.
\end{equation}
 \section{Illustrative Examples }
 
 \textbf{Example 1}. A particle free to travel in a small space surrounded by impenetrable barriers defines the infinite potential well. 
In this model, potential energy is given as
\begin{equation}
V_\alpha(x)=\begin{cases}
v_\alpha(x) & \quad \text{if} \quad 0<x<L \\
\infty & \quad \text{if} \quad x<0,x>L,
\end{cases}
\end{equation}
where assume the potential $ v_\alpha(x)$ to be slowly varying . To calculate the fractional energy using eq,(\ref{psi sin}), with apply the boundary condition of $x=0 \to \psi(0)=0, x=L \to \psi(L)=0 $, where in $x=0 \to \psi(0)=0 \to C_1 \neq 0, C_2=0$, thus, using condition $ x=L \to \psi(L)=0 $, we obtain 
\begin{equation}
\phi(L)=\int_0^L p^\alpha (x) d^\alpha x=n\pi \hbar^{\alpha}_\alpha, \quad n=1,2,3, \dots 
\end{equation}
In special case $V_\alpha(x)=0$, we have
\begin{equation}
\phi(L)=\int_0^L \sqrt{2m^\alpha E^\alpha } x^{\alpha-1} dx = n\pi \hbar^{\alpha}_\alpha,
\end{equation}
thus, we obtain the conformable energy levels
\begin{equation}
E^\alpha=\frac{n^2 \alpha^2 \pi^2 \hbar^{2\alpha}_\alpha}{2m^\alpha L^{2\alpha}}, \quad n=1,2,3, \dots 
\end{equation}
The conformable wave function is obtained as form
\begin{equation}
\psi_\alpha(x)=C_1 \sqrt{\frac{L^\alpha}{n \alpha \pi \hbar^{\alpha}_\alpha } }\sin\left(\frac{n\pi x^\alpha }{L^\alpha}\right),
\end{equation}
where $C_1 = \frac{\alpha \sqrt{2n\pi \hbar^{\alpha}_\alpha}}{L^\alpha}$ and this equation becomes 
\begin{equation}
\psi_\alpha(x)=\sqrt{\frac{2\alpha}{L^\alpha }}\sin\left(\frac{n\pi x^\alpha }{L^\alpha}\right).
\end{equation}
The Schrodinger for  equation infinite potential well is solved by Chung et.al.\cite{chung2020effect} and they obtained the  same result.\\
 \textbf{Example 2}. We consider here the damped harmonic oscillator. The fractional Hamiltonian of damping harmonic oscillator for Bateman system can be written as \cite{serhan2018quantization}
\begin{equation}
\begin{aligned}
& H_\alpha= \frac{(P^\alpha_y)^2}{2m^\alpha}+\frac{m^\alpha}{2}\left(\omega^{2\alpha}-\frac{\lambda^2}{4}\right)y^{2\alpha} \\ & \quad \label{bat} 
= -\frac{\hbar^{2\alpha}_\alpha}{2m^\alpha}D^\alpha_y D^\alpha_y + \frac{m^\alpha}{2}\left(\omega^{2\alpha}-\frac{\lambda^2}{4}\right)y^{2\alpha},
\end{aligned}
\end{equation}
$\lambda$ is the damping constant. We assume the damped part is perturbed. To calculate the conformable energy for damping harmonic oscillator (Batman system) using   eq.(\ref{phi fun}), we have
	\begin{equation}
	\label{main relation}
		\phi=\frac{\sqrt{2m^\alpha}}{\hbar^{\alpha}_\alpha}\int_{y^{\alpha}_1}^{y^{\alpha}_2} \sqrt{ E^\alpha-\frac{m^\alpha}{2}\left(\omega^{2\alpha}-\frac{\lambda^2}{4}\right) y^{2\alpha}} d^\alpha y.
\end{equation}
The inflection point when $P^\alpha_y=0 \rightarrow E^\alpha= \frac{m^\alpha}{2}(\omega^{2\alpha}-\frac{\lambda^2}{4})y^{2\alpha} $. So, we get 
\begin{equation}
\label{value y}
y^{\alpha}_1=- \sqrt{\frac{E^\alpha}{\frac{m^\alpha}{2}\left(\omega^{2\alpha}-\frac{\lambda^2}{4}\right)}} \quad , y^{\alpha}_2= \sqrt{\frac{E^\alpha}{\frac{m^\alpha}{2}\left(\omega^{2\alpha}-\frac{\lambda^2}{4}\right)}}.
\end{equation}
Thus, we find $ y^{\alpha}_1=-y^{\alpha}_2$, and plug in eq.(\ref{main relation}), we have 
\begin{equation}
\label{int1}
\phi=\frac{m^\alpha}{\hbar^{\alpha}_\alpha}\sqrt{\left(\omega^{2\alpha}-\frac{\lambda^2}{4}\right)}\int_{-y^{\alpha}_2}^{y^{\alpha}_2}\sqrt{\frac{E^\alpha}{\frac{m^\alpha}{2}\left(\omega^{2\alpha}-\frac{\lambda^2}{4}\right)}-y^{2\alpha}}  y^{\alpha-1} dy
\end{equation}
The solution for this integral is given by
\begin{equation}
\int_{-y^{\alpha}_2}^{y^{\alpha}_2}\sqrt{\frac{E^\alpha}{\frac{m^\alpha}{2}\left(\omega^{2\alpha}-\frac{\lambda^2}{4}\right)}-y^{2\alpha}}  y^{\alpha-1} dy=\frac{y^{2\alpha}_2}{\alpha}\frac{\pi}{2}.
\end{equation}
So, substituting in eq.\eqref{int1}, we have 
\begin{equation}
\label{phi}
\phi=\frac{m^\alpha}{\hbar^{\alpha}_\alpha}\sqrt{\omega^{2\alpha}-\frac{\lambda^2}{4}}\frac{y^{2\alpha}_2}{\alpha}\frac{\pi}{2},
\end{equation}
using connection formula condition \cite{zettili2003quantum},  
\begin{equation}
\phi=\left(n+\frac{1}{2}\right)\pi.
\end{equation}
And, substituting in eq.(\ref{phi}), we have 
\begin{equation}
\label{y}
 y^{2\alpha}_2=\frac{2\alpha \hbar^{\alpha}_\alpha}{m^\alpha \sqrt{\omega^{2\alpha}-\frac{\lambda^2}{4}}}\left(n+\frac{1}{2}\right),
\end{equation}
from eq \eqref{value y} we find $y^{2\alpha}_2= \frac{E^\alpha}{\frac{m^\alpha}{2}(\omega^{2\alpha}-\frac{\lambda^2}{4})} $ plug in \eqref{y} we get 
\begin{equation}
\label{energy}
E^\alpha=\hbar^{\alpha}_\alpha \alpha \sqrt{\omega^{2\alpha }-\frac{\lambda^2}{4}}\left(n+\frac{1}{2} \right).
\end{equation}
It is in exact agreement with \cite{serhan2018quantization}, when $\alpha=1$.
\\\\
\textbf{Example 3}. We consider the potential of alpha particle 
\begin{equation}
\label{alpha potential}
V_\alpha(r) =\begin{cases}
\frac{A^\alpha}{\alpha r^\alpha} & \quad \text{if} \quad r^\alpha_1 < r^\alpha, \\
0 & \quad \text{if} \quad r^\alpha_1 > r^\alpha, 
\end{cases} 
\end{equation}
where $ A^\alpha=\frac{2ze^2}{4\pi \epsilon_0} $.\clearpage
\begin{figure}[hbt]
\begin{center}
\scalebox{0.7}{\includegraphics{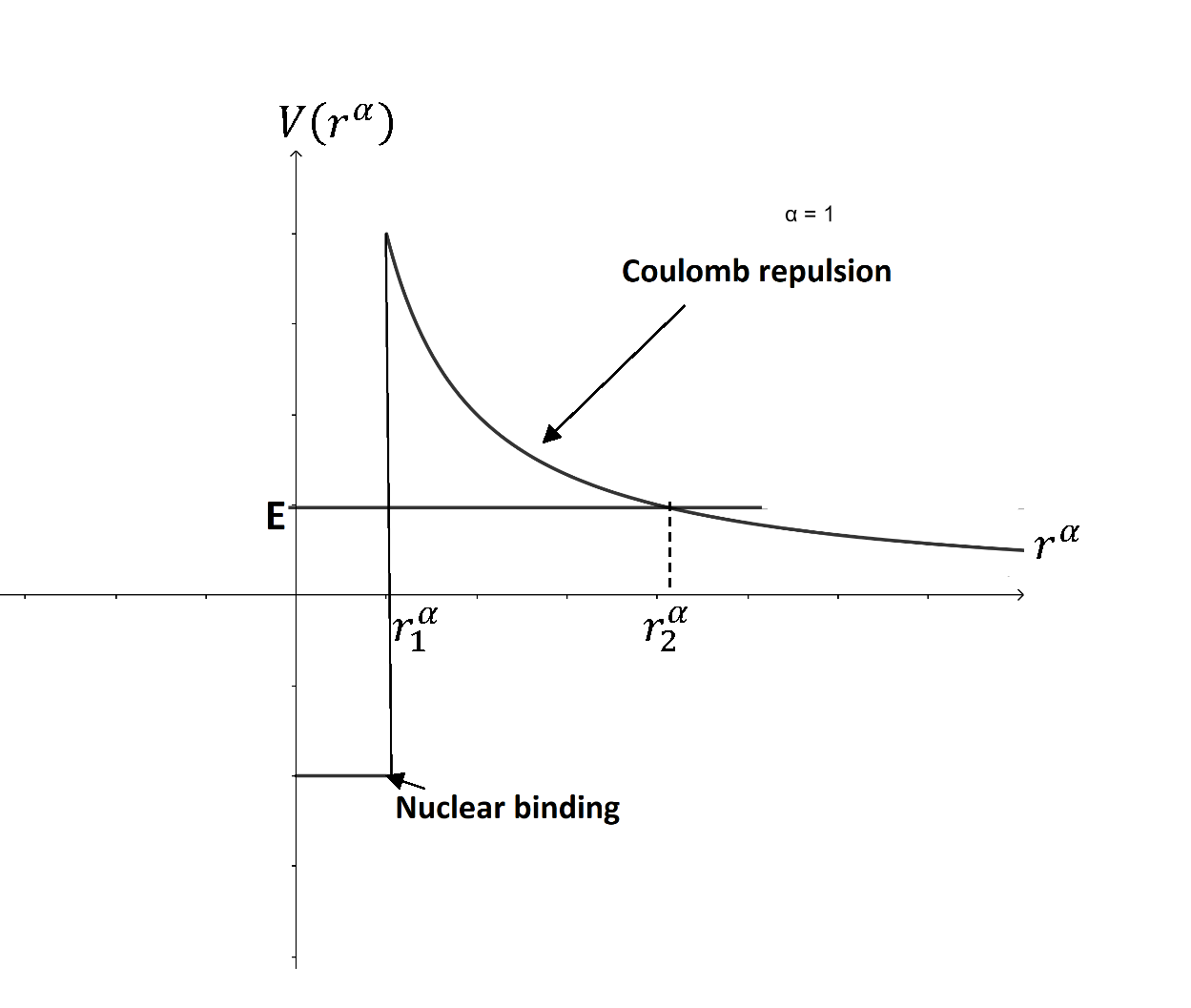}}
\caption{\label{fig:lj}Potential model for $\alpha-$particle's, in case $\alpha=1$}
\end{center}
\end{figure}
 Here $r_2^\alpha$ denotes the point where the energy equals the potential which w may call it the conformable turning point. Thus, the conformable energy of alpha particle is given by
\begin{equation}
\label{alpha energy}
E^\alpha = \frac{A^\alpha}{\alpha r_2^\alpha},
\end{equation}
 using eq.(\ref{gamow}) to calculate transmission probability
 
 \begin{equation} \begin{aligned}
    & \gamma_\alpha = \frac{1}{\hbar^{\alpha}_\alpha}\int_{r_1^\alpha}^{r_2^\alpha} \frac{\sqrt{2m^\alpha(\frac{A^\alpha}{\alpha r^\alpha}-E^\alpha)}}{r^{1-\alpha}} dr\\
&\quad = \frac{\sqrt{2m^\alpha E^\alpha }}{\hbar^{\alpha}_\alpha}\int_{r_1^\alpha}^{r_2^\alpha}\sqrt{(\frac{r_2^\alpha}{ r^\alpha}-1)}r^{\alpha-1} dr \\
&\quad \label{result} 
 =\frac{\sqrt{2m^\alpha E^\alpha }}{\hbar^{\alpha}_\alpha}[r_2^\alpha\arccos{\sqrt{\frac{r_1^\alpha}{r_2^\alpha}}}-\sqrt{r_1^\alpha(r_2^\alpha-r_1^\alpha)} ].
 \end{aligned}
\end{equation}
In case $r_1^\alpha \ll r_2^\alpha \to \arccos{\sqrt{\frac{r_1^\alpha}{r_2^\alpha}}} \approx \frac{\pi}{2}-\sqrt{\frac{r_1^\alpha}{r_2^\alpha}}$ and $r_2^\alpha-r_1^\alpha \approx r_2^\alpha $, so we have 
\begin{equation}
\gamma_\alpha = \frac{\sqrt{2m^\alpha E^\alpha }}{\hbar^{\alpha}_\alpha}[r_2^\alpha\frac{\pi}{2}-2\sqrt{r_1^\alpha r_2^\alpha} ].
\end{equation}
Substituting value of $r_2^\alpha$ from eq.(\ref{alpha energy}), we have
\begin{equation}
\gamma_\alpha = K_1^\alpha (\pi \sqrt{2})^{1-\alpha} \frac{z^\alpha}{\alpha\sqrt{E^\alpha}}- K_2^\alpha 4^{1-\alpha} \sqrt{\frac{r_1^\alpha z^\alpha}{\alpha}},
\end{equation}
where $K_1=\frac{e^2 \pi \sqrt{2m}}{4\pi \epsilon_0 \hbar}=1.986 MeV$ and $K_2=(\frac{e^2 m }{4\pi \epsilon_0 })^{\frac{1}{2}} \frac{4}{\hbar}=1.485 MeV$ \cite{griffithsIntroductionQuantumMechanics2018}. Thus, after substituting this equation in eq.(\ref{tran}), we have
\begin{equation}
T=\exp{-2\left( K_1^\alpha (\pi \sqrt{2})^{1-\alpha} \frac{z^\alpha}{\alpha\sqrt{E^\alpha}}- K_2^\alpha 4^{1-\alpha} \sqrt{\frac{r_1^\alpha z^\alpha}{\alpha}}\right)}.
\end{equation}
It is in exact agreement with \cite{griffithsIntroductionQuantumMechanics2018}, when $\alpha = 1$.
\section{\textbf{Summary and conclusion}}
The WKB approximation is extended to be applicable for $\alpha-$Hamiltonian in the conformable  form. The conformable derivative of $\alpha$ fractional order is used and applied  to three illustrative examples. Quantities were obtained in the conformable form so that the corresponding the traditional forms are recovered when  $\alpha = 1$
\bibliography{ref-WKB}
\bibliographystyle{IEEEtran}
\end{document}